\documentclass[twocolumn]{aastex631}

\usepackage{apjfonts}

\usepackage{graphicx}
\usepackage{color}

\usepackage{amsbsy}

\usepackage{mathrsfs}

\usepackage{mathpazo,bm}

\usepackage{amsmath}

\usepackage{xfrac}
\usepackage{url}
\usepackage{hyperref}
\usepackage{enumitem}
\usepackage{natbib}

\newlength{\figspace}
\newcommand{\beq}{\begin{equation}}
\newcommand{\eeq}{\end{equation}}
\newcommand{\beqn}{\begin{eqnarray}}
\newcommand{\eeqn}{\end{eqnarray}}

\newcommand{\St}{\mathrm{St}}

\newcommand{\xtimes}[2]{#1\times{10^{#2}}}

\newcommand{\Laplace}{\nabla^2}

\newcommand{\Eq}[1]{Eq.~(\ref{#1})}

\newcommand{\eq}[1]{\Eq{#1}}
\newcommand{\eqp}[1]{(Eq.~\ref{#1})}

\newcommand{\Fig}[1]{Fig.~\ref{#1}}

\renewcommand{\fig}[1]{\Fig{#1}}

\definecolor{brown}{rgb}{0.42,0.24,0.07}
\definecolor{darkgreen}{rgb}{0.0,0.6,0.00}
\definecolor{purple}{rgb}{0.7,0.0,0.7}
\definecolor{black}{rgb}{0.0,0.0,0.0}

\def\apj{\rm ApJ}
\def\apjl{\rm ApJL}

\def\mnras{\rm MNRAS}
\def\nat{\rm Nature}

\def\aap{\rm A\&A}

\def\icarus{\rm Icarus}

\usepackage[encapsulated]{CJK}
\newcommand{\cntext}[1]{\begin{CJK}{UTF8}{bkai}#1\ignorespacesafterend\end{CJK}} 

\received{February 22th, 2024}
\accepted{June 24th, 2024}
\submitjournal{ApJ}

\shorttitle{Vortex trapping}
\shortauthors{Lyra et al.}

\begin{document}

\title{Rapid protoplanet formation in vortices: three-dimensional local simulations with selfgravity}

\correspondingauthor{Wladimir Lyra}
\email{wlyra@nmsu.edu}

\author[0000-0002-3768-7542]{Wladimir Lyra}
\affiliation{New Mexico State University, Department of Astronomy, PO Box 30001 MSC 4500, Las Cruces, NM 88001, USA}

\author[0000-0003-2589-5034]{Chao-Chin Yang (\cntext{楊朝欽})}
\affiliation{Department of Physics and Astronomy, The University of
  Alabama, Box~870324, Tuscaloosa, AL~35487-0324, USA}

\author[0000-0002-3771-8054]{Jacob B. Simon}
\affiliation{Department of Physics and Astronomy, Iowa State University, Ames, IA, 50010, USA}

\author[0000-0001-5372-4254]{Orkan M. Umurhan}
\affiliation{NASA Ames Research Center, Space Sciences Division,
  Planetary Sciences Branch, Moffatt Field, CA 94035, USA}
\affiliation{SETI, Carl Sagan Center, 190 Bernardo Way, Mountain View, CA 94043, USA}

\author[0000-0002-3644-8726]{Andrew N. Youdin}
\affiliation{Department of Astronomy and Steward Observatory, University of Arizona, Tucson, Arizona 85721, USA} 
\affiliation{The Lunar and Planetary Laboratory, University of Arizona}

\begin{abstract}

  Disk vortices, seen in numerical simulations of protoplanetary disks and found
  observationally in ALMA and VLA images of these objects, are
  promising sites for planet formation given their pebble trapping
  abilities. Previous works have shown strong concentration of
  pebbles in vortices, but gravitational collapse has only been shown in low-resolution, two-dimensional, global
  models. In this letter, we aim to study the pebble concentration and gravitational collapse of pebble clouds in vortices
  via high-resolution, three-dimensional, local models. We performed simulations of the dynamics of gas and solids in a local
  shearing box where the gas is subject to convective overstability, generating a persistent giant vortex. We find that the
  vortex produces objects of Moon and Mars mass, with 
  mass function of power law $d\ln N/d\ln M=-1.6\pm 0.3$. The protoplanets grow rapidly,  doubling in mass in about 5 orbits,
  following pebble accretion rates. The
  mass range and mass doubling rate are in broad agreement with previous
  low resolution global models. We conclude that Mars-mass planetary embryos are the
  natural outcome of planet formation inside the disk vortices seen in millimeter and radio images of protoplanetary disks. 

\end{abstract}


\section{Introduction} 
\label{sect:introduction}

Anticyclonic vortices in circumstellar disks have long been heralded as promising sites for planet
formation, ever since they were independently suggested as such by
\cite{BargeSommeria95}, \cite{AdamsWatkins95}, and
\cite{Tanga+96}. While cyclones are low pressure regions, in a
disk the Keplerian shear ensures that only anti-cyclones survive. 
These anti-cyclones are pressure maxima, making them natural traps of dust pebbles. 
A gas parcel in a vortex is in geostrophic equilibrium 
between the negative pressure gradient, the Coriolis force, and the centrifugal force. 
In essence, the vortex behaves like a miniature disk, with the Coriolis force playing 
the role that stellar gravity does in the larger circumstellar disk. The pebble dynamics 
is likewise analogous. Once inside the vortex, a pebble orbits the
vortex center due to the Coriolis force; yet, the pebble continuously
loses angular momentum due to the drag force. As a result, pebbles
drift steadily toward the center of the vortex, where they accumulate.

It was shown by \cite{Lyra+08b,Lyra+09a,Lyra+09b} that in the limit where turbulence in the vortex core is absent, the concentration of pebbles easily reaches the conditions necessary
to gravitationally collapse them into planets. These simulations,
albeit 2D, show that disk vortices trap an ensemble of pebbles of total mass around
the mass of Mars. \cite{Meheut+12a} independently find similar results
with 3D simulations. These results, however, were obtained with global
simulations and thus of low resolution in the region of interest; most importantly, the
turbulence in the vortex core could not be realized. Fluids in uniform
rotation support a spectrum of stable inertial waves
\citep[c.f.][]{Chandrasekhar61}. Strain is introduced when the motion
passes from circular to elliptical, and some three-dimensional modes
find resonance with the underlying strain field. The resulting
elliptical instability
\citep{Bayly86,Pierrehumbert86,Kerswell02,LesurPapaloizou09} easily
breaks the coherence of the vortex, feeding on its kinetic energy,
that then cascades and dissipates. If vorticity is supplied in
  the integral scale, so that a steady state of ``elliptic
  turbulence'' is achieved, the vortex core develops rms velocities in the range 5\%-10\%
of the sound speed \citep{LesurPapaloizou10,LyraKlahr11}.

The impact of this vortex core turbulence in pebble dynamics was
worked out analytically by \citet{KlahrHenning97} and \citet{LyraLin13}, showing
that a steady state is reached where the inward drift of pebbles is balanced by turbulent
diffusion. Simulations by \citet{Lyra+18} and \citet{Raettig+21} detail how this
turbulent diffusion happens hydrodynamically, showing that the pebbles disrupt the
vortex around the midplane but that the coherence of the
three-dimensional vortex column is maintained.

These advances in the understanding of vortices in the past decade
still left unanswered what the mass distribution of planets formed in
vortices is, a quantity (the mass function) that has been measured in
streaming instability calculations (though for much smaller mass objects; e.g., \citealt{Johansen+15,Simon+16,Schafer+17,Li+19,Gole+20}). High concentrations of pebbles
with dust-to-gas ratio above unity were observed in the simulations of
\cite{Raettig+21}, supporting the notion that the formation of planetary bodies was a likely
outcome of the trapping. Indeed, some of the enhancements were above
Hill density, and would have collapsed gravitationally had
selfgravity been included. Here we use the same model as
\cite{Raettig+21}, now including self-gravity, to determine the
mass distribution of protoplanets formed in baroclinic vortices. 

This letter is structured as follows. We introduce the equations of motion in Sect 2, followed by the results in Sect 3. We conclude the letter with a discussion of the results in Sect 4. 


\section{Model}
\label{sect:physicalbackground}

We use a baroclinic shearing box with the pressure gradient linearized, and an
optically thin cooling law. The gas model is identical to the one used in
\cite{LyraKlahr11,Lyra14,Raettig+13,Raettig+15,Lyra+18}, and
\citet{Raettig+21}, except that here we also include self-gravity of
the particles. We solve the Poisson equation  

\begin{equation}
  \Laplace\varPhi = 4\pi G \rho_d\label{eq:poisson}
\end{equation}

\noindent via fast Fourier transforms, as in \citet{Johansen+07}. Here
$\varPhi$ is the selfgravitational potential, $\rho_d$ is
the dust density, and $G$ is the gravitational constant. The strength of selfgravity is set by the dimensionless parameter

\begin{equation}
  \tilde{G} =\frac{4\pi G \rho_0}{\varOmega^2},
  \label{eq:Gtilde}
\end{equation}

\noindent where $\rho_0$ is the reference density and $\varOmega$ is the
Keplerian frequency. The parameter $\tilde{G}$ is related to the Toomre $Q$ parameter (for the gas) by $Q=\tilde{G}^{-1}\sqrt{8/\pi}$.

As particles concentrate, a gravitationally bound clump is replaced by a sink particle when the
particle density increases above the Hill density \citep{Johansen+15}, 

\begin{equation}
\rho_R \equiv \frac{9\varOmega^2}{4\pi G},
\end{equation}

\noindent  meaning that self-gravity overcomes tidal forces. This criterion does not include the effect of diffusion, which increases the density needed
to collapse \citep{KlahrSchreiber20,KlahrSchreiber21}. As the simulation progresses, sink
particles are free to grow in mass by accreting pebbles and other protoplanets.  

We quote the masses in units of the Toomre mass \citep[e.g.][]{Abod+19,Li+19},
which is the mass enclosed in a circular area of diameter equal to the
Toomre wavelength $\lambda_G$ and column density of pebbles $\varSigma_p$, 

\beq
M_G = \pi \left(\frac{\lambda_G}{2}\right)^2 \varSigma_p,
\eeq

\noindent where the Toomre wavelength is 

\beq
\lambda_G = \frac{4\pi^2 G \varSigma}{\varOmega^2},
\eeq

\noindent and $\varSigma$ is the gas column density.

The box length is $4H \times 16H \times 2H$, where $H\equiv c_{s0}/\varOmega$ is
the scale height, and $c_{s0}$ is the reference sound speed. The box is vertically
unstratified, with periodic boundary conditions in $y$ and $z$, and
shear-periodic in $x$.  The simulations were done with the {\sc Pencil
  Code} \citep{BrandenburgDobler02,BrandenburgDobler10,Brandenburg+21}
and carried out at multiple
resolutions as described in more detail below. 

The global dust-to-gas ratio is set to $\xtimes{3}{-2}$, and the Stokes number is
0.3, well in the regime of strong clumping
\citep{Carrera+15,Yang+17,LiYoudin21} though we do not
  expect streaming instability, given our resolution. We use code units such that

\beq
c_p =  \varOmega = \rho_0  =1,
\eeq

Here $c_p$ is the heat capacity at constant pressure, and $\rho_0$ is
the reference volume density. Other parameters are the adiabatic index
$\gamma=1.4$, and $\tilde{G}=0.1$. This value of $\tilde{G}$ corresponds to $Q\approx 16$, so the gas
selfgravity is negligible. We also define the aspect ratio
  $h\equiv H/r$ = 0.1 and finally, we set $\Pi\equiv\Delta v / c_{s0} = 0.125$, where $\Delta v$
is the velocity reduction on the gas, with respect to Keplerian, due
to the global pressure gradient. This value is justified given $\Delta v =
  v_K  \mathcal{O}(h^2)$, i.e., the Keplerian velocity $v_k$ scaled
  by a factor of second order in aspect ratio, so the value of $\Pi$ is of
  first order in aspect ratio,  i.e., $\Pi  = \mathcal{O}(h)$.

Conversion from code units to physical units necessitates choosing a
box location and stellar mass, which sets the physical value of
$\varOmega$. For a box orbiting a solar mass at 20\,AU, the
reference volume density at the midplane is then given by
\eq{eq:Gtilde}, yielding $\rho_g \approx \xtimes{6}{-13}$
g\,cm$^{-3}$. The column density is $\Sigma \approx 44$
g\,cm$^{-2}$. The box has 12.75 Earth masses of dust. The Toomre mass is $M_G \approx \xtimes{8.7}{-5}
M_\oplus$. Assuming $\rho_\bullet = 3$\,g\,cm$^{-3}$ for the grains,
the Stokes number $\St=0.3$ implies grain radius $a_\bullet = 1.7$ cm.

\begin{figure*}
  \begin{center}
    \resizebox{\textwidth}{!}{\includegraphics{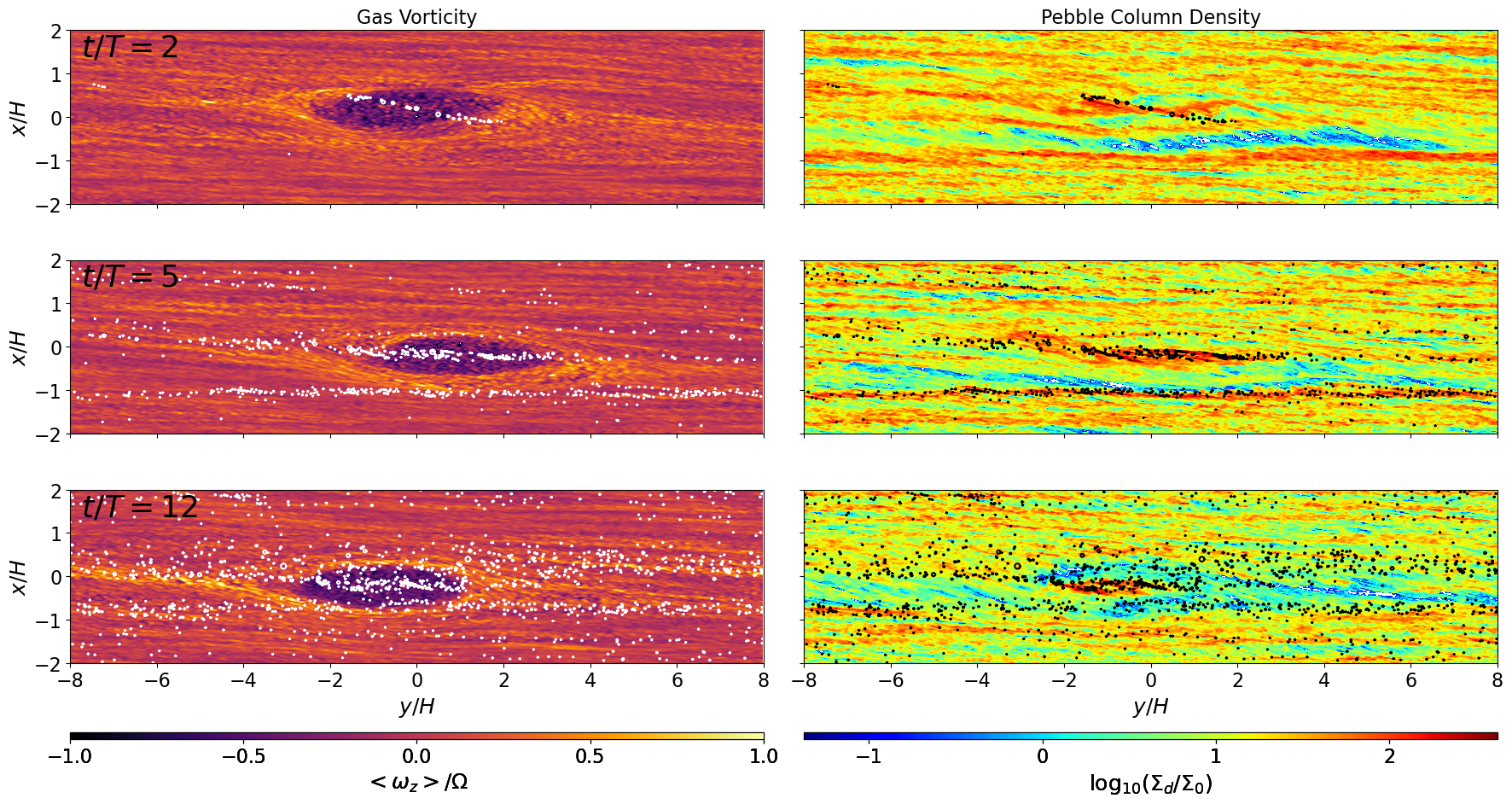}}
  \end{center}
\caption{Distribution of protoplanets formed in the higher resolution model
  (512$^3$ mesh resolution, $7.5\times 10^6$ superparticles). Left panels
  show the vertically averaged vertical vorticity, right panels show
  the vertically integrated (column) density of pebbles. The upper
  panels correspond to 2 orbits after the
  insertion of particles, middle panel after 5, and lower panel after
  12 orbits. The protoplanets are overlaid as white circles in the
  left panels, black circles in the right panels. The radius of the
  circle marks the Hill radius of the protoplanet. Protoplanets are formed inside the vortex, which
  continuously spawns planetary mass objects.}
\label{fig:vorticitypebbles}
\end{figure*}

\begin{figure*}
  \begin{center}
    \resizebox{\textwidth}{!}{\includegraphics{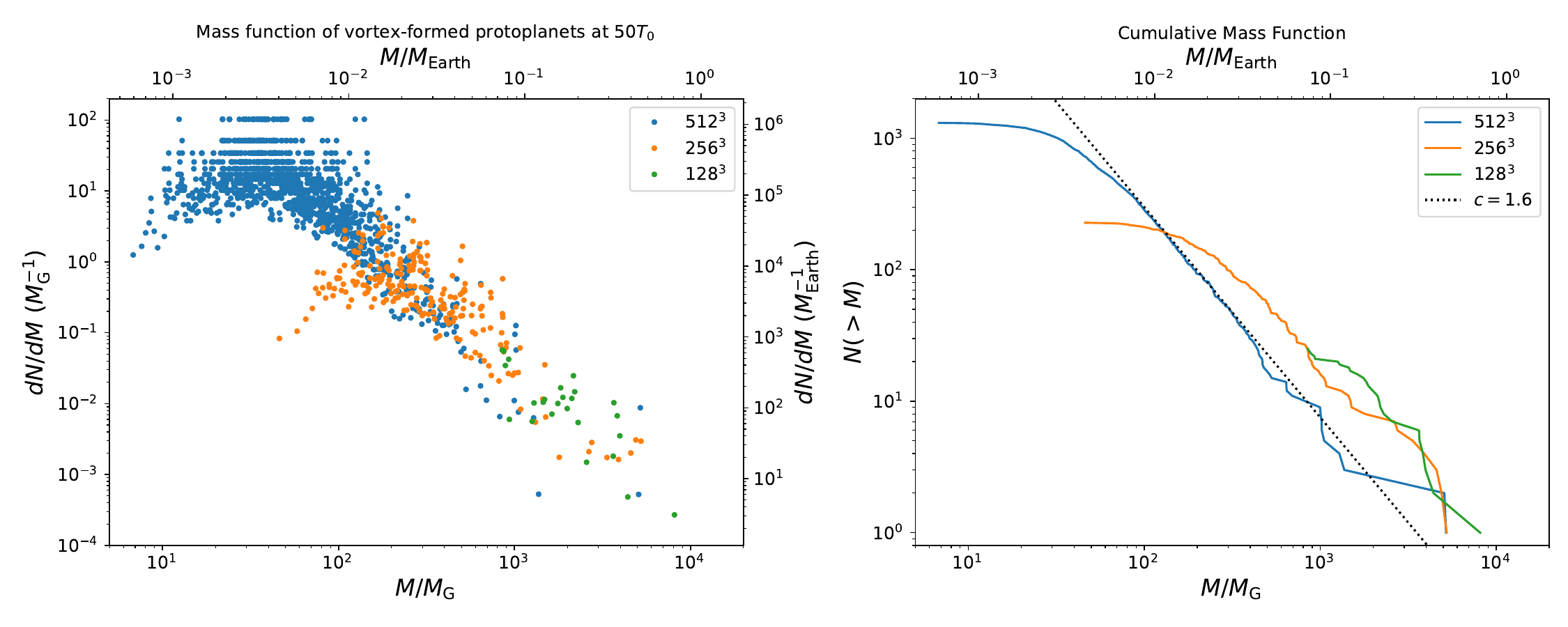}}
  \end{center}
\caption{{\it Left:} Differential Mass Function, 50 orbits after particle
  insertion, as a function of resolution. The simulations
converge at high mass, but as we increase the resolution,
more small mass objects are formed, with no convergence in
sight. The horizontal stripes seen from our highest $512^3$ resolution 
is due to the mass discretization of the super-particles; see \eq{eq:mass-function}. {\it Right:} Cumulative mass function, 50 orbits after particle
  insertion, as a function of resolution.}
\label{fig:imf-res}
\end{figure*}

\begin{figure}
  \begin{center}
    \resizebox{\columnwidth}{!}{\includegraphics{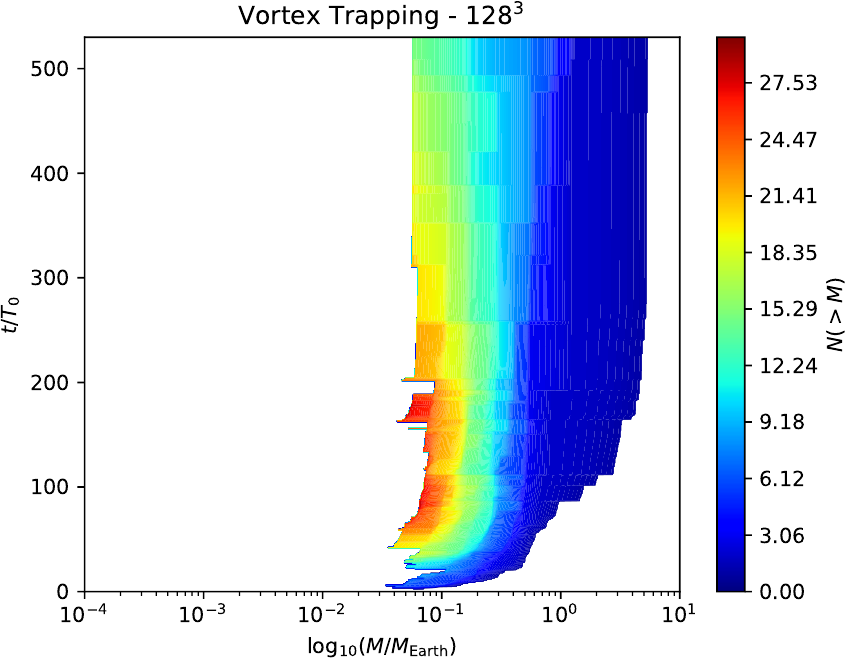}}
    \resizebox{\columnwidth}{!}{\includegraphics{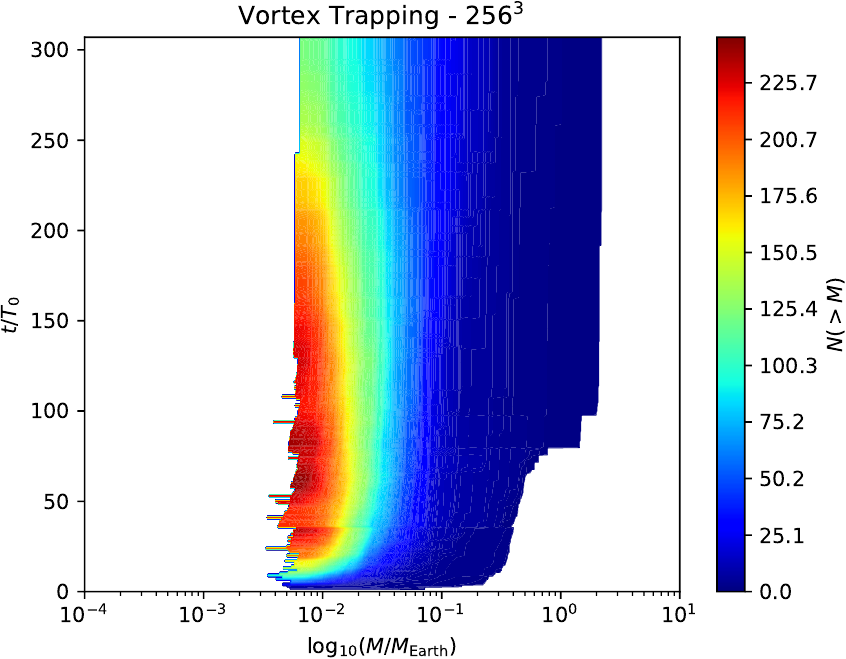}}
    \resizebox{\columnwidth}{!}{\includegraphics{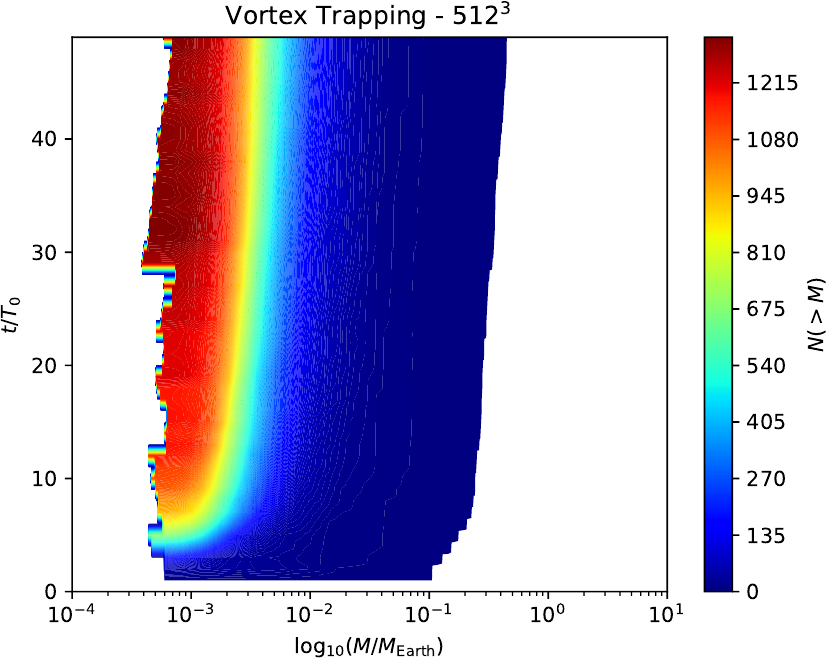}}
  \end{center}
\caption{Evolution of the cumulative mass distribution, as a function
  of resolution. Notice the different final times. The mass
  distribution did not converge in time or resolution.}
\label{fig:cumm-time}
\end{figure}

\begin{figure}
  \begin{center}
    \resizebox{\columnwidth}{!}{\includegraphics{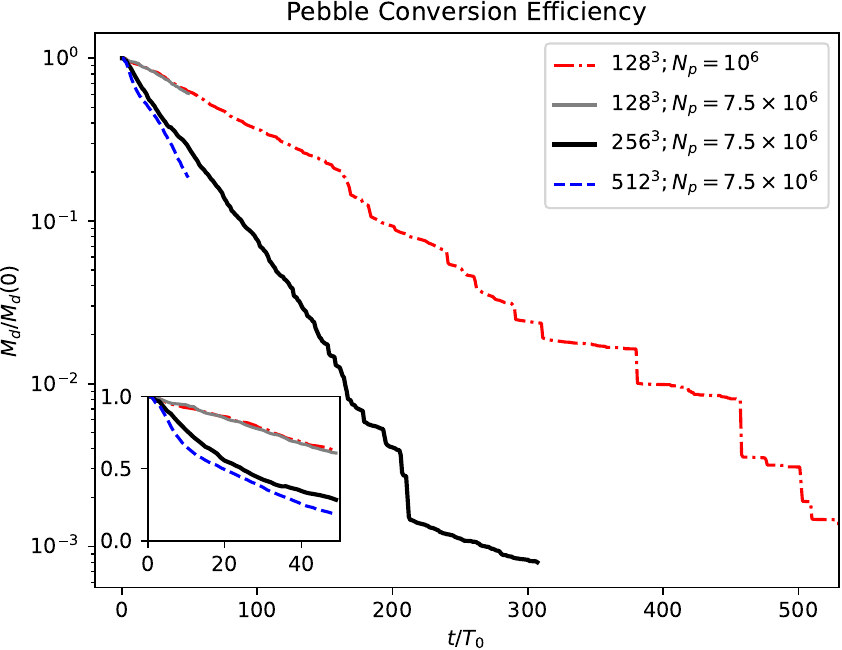}}
 \end{center}\caption{Efficiency of conversion of pebbles into
   protoplanets. The y-axis in the inset
   is shown in linear scale.
   Seen from the $128^3$ at
   different particle resolution, that quantity has
   converged. Convergence is not seen with mesh resolution, although
   $256^3$ to $512^3$ has not changed the rate as much as from $128^3$
   to $256^3$. At $256^3$ resolution the available dust density
   drops an order of magnitude in 100 orbits.}
\label{fig:conversion}
\end{figure}

\begin{figure*} 
  \begin{center}
    \resizebox{\textwidth}{!}{\includegraphics{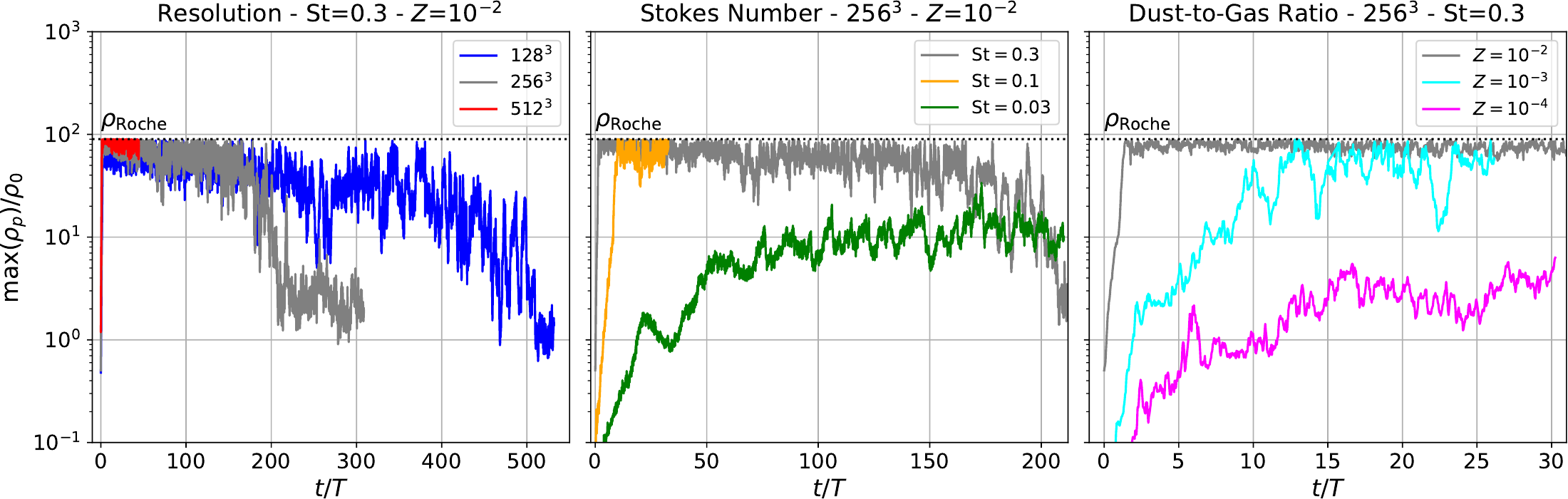}}
  \end{center}
  \caption{Time evolution of the maximum pebble density as a
      function of resolution (left), Stokes number (middle)
      and dust-to-gas ratio (right). The Hill density for this simulation
    is $\rho_d = 90$ in code units. When a clump of pebbles achieves
    this density, a sink particle replaces it. }
\label{fig:pebble_evolution}
\end{figure*}

\begin{figure} 
  \begin{center}
    \resizebox{\columnwidth}{!}{\includegraphics{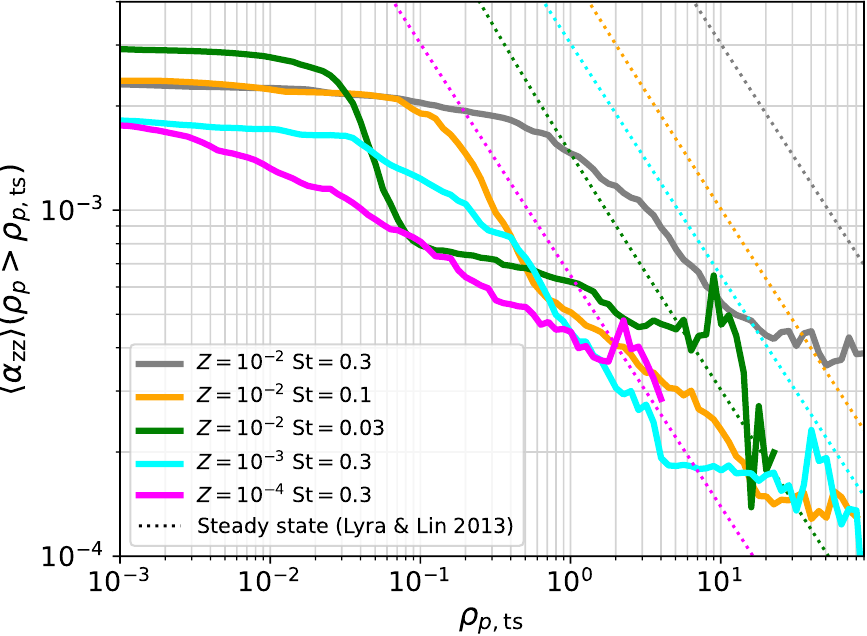}}
  \end{center}
  \caption{Density-diffusion relation for the $256^3$ resolution runs shown in
      \fig{fig:pebble_evolution}. The $x$-axis shows a pebble density
      threshold, and the $y$-axis the diffusion computed as an average
      of all mesh cells with pebble density above the threshold, in
      space and time (see text). The dotted lines show the
      steady-state solution of \citet{LyraLin13} for pebbles trapped in a
      vortex considering stellar vertical gravity, drag force, and diffusion. The runs
      that achieve collapse (Hill density $\rho_p = 90$) have
      diffusion under the steady-state prediction. The runs that do
      not (green line and magenta line) have maximum density
      consistent with the analytical solution.}
\label{fig:alpha-rhop}
\end{figure}

The gas model is identical to \cite{Lyra14,Lyra+18} and
\cite{Raettig+21}, and the reader is referred to these papers for an
exhaustive description. For the fiducial model we use resolution $256^3$ (we also ran 128$^3$ and 512$^3$). The cells
are non-cubic; the equal number of cells in different directions is
needed for the Poisson solver. The box is seeded with Gaussian noise
in the velocity at the $10^{-3} c_s$ level. The cooling time is $\tau
= \gamma^{-1}$, to maximize the growth rate of the convective
overstability. The simulation is ran for 400 orbits, until the
convective overstability saturates and a single large vortex dominates
the box. We emphasize that these thermodynamical conditions are
  unrealistic, but our goal in using the convective overstability is
  simply to produce a vortex in a robust and generic way. More
  realistic thermodynamics would simply slow down the evolution to a
  similar saturated state.

After the vortex saturates, $7.5\times 10^6$ super-particles are
introduced. A super-particle represents a swarm of pebbles sharing
the same position, velocity, and aerodynamical properties. At this resolution, the mass of an
individual super-particle is $10^{22}$ g ($10^{-4}$ lunar masses, or $10^{-2}$
Ceres mass).

\section{Results}

\Fig{fig:vorticitypebbles} overplots the gravitationally bound objects formed (black and white symbols) against the $z-$averaged gas vertical vorticity (left panel) and the pebble column
density (right panels). The upper plots are from a snapshot at 2 orbits
after inclusion of particles, the middle one at 5, and the lower at 12
orbits. We see at 2 orbits that indeed the formation of the
gravitationally bound clumps happens inside the vortex. The vortex is colocated with
an extended region of high pebble density, which spawns the
clumps. The middle panel shows that the objects do not
stay in the vortex because, unlike the pebbles, they do not feel the gas
drag. The vortex indeed functions as a ``planet formation
factory'', turning pebbles into planets, that then leave the
vortex, which in turn forms more planets. A ridge of high
  pebble density, colocated with bound objects, is seen at
  $x\approx-0.1H$. The lower panel shows
the spatial distribution of the protoplanets at 12 orbits. 

\subsection{The vortex spawns protoplanets}

The typical mass of the objects formed is between the mass of the Moon
and Mars, making them not planetesimals, but {\it protoplanets}. At two orbits in 256$^3$ resolution, eight protoplanets are already formed, the smallest
mass at 0.4 lunar masses, and the biggest at 5 lunar masses. Given the
particle resolution ($10^{22}$ g), these masses are resolved,
particle-wise. At the grid level, though, the mass at the time of collapse is
close to the Hill density. A cell volume at 256$^3$ resolution is $\Delta
x \Delta y \Delta z = 2\times 10^{35}$ cm$^3$. The Hill density
($\approx 5 \times 10^{-11}$) filling this cell corresponds to a mass
of $10^{25}$ g, or 0.15 lunar mass, close to the lowest mass formed. 

We run a resolution study to check if the initial mass function changes with grid resolution. 
\fig{fig:imf-res} shows the mass functions, differential (left panel) and
cumulative (right panel) for different resolutions, 128$^3$, 256$^3$, and 512$^3$, taken at 50 orbits
after particle insertion. The differential mass function is calculated
as

\begin{equation}
\left.\frac{dN}{dM}\right\vert_i = \frac{2}{M_{i+1}-M_{i-1}}
\label{eq:mass-function}
 \end{equation}

\noindent i.e. the second-order accurate centered first derivative. This form is used
 except for the least massive and most massive objects, where $i-1$ and $i+1$ do
 not exist, respectively, in which case the first-order accurate
 forward and backward derivative is used, respectively.  

 As resolution increases, more protoplanets
 of low mass are formed, the lowest being $5\times 10^{-4} M_\oplus$
 ($\approx$ 3 Ceres masses), with no convergence seen at low
   masses as we increase the resolution. 

 The cumulative mass function shows that a power law of -1.6 follows in
 the resolved domain, flattening toward the bottom quarter of the
 logarithmic domain in mass. 

In \fig{fig:cumm-time}  we plot the time evolution of the cumulative
mass distribution, to check for convergence in time, i.e. if the
distribution has reached a steady state. Snapshots at every orbit are
taken, and the mass function (which is continuous) is interpolated to
a range from $10^{-4}$ to $10 M_\oplus$, with 1000 points, log-spaced. We see that after planet formation starts, the number of objects in all mass bins
increase steadily. The formation of lower mass objects at increasing
resolution is also clear. Some pixelization at low and high masses is seen as a result of
the interpolation procedure. At later times, seen in the $128^3$ and
$256^3$ runs (notice the different final times), growth stalls 
at the high-mass end. The timescale of the hydrodynamical simulation does not allow for 
convergence in the high-mass end, as binary collisions become the main driver of mass growth, which 
happen much longer timescales \citep{Sandor+11}.

To check for particle resolution, we run a simulation with only
  $10^6$ particles at $128^3$ resolution, to compare with the ``fiducial''
  particle resolution of $7.5\times 10^6$. \fig{fig:conversion} shows that the 
  efficiency of pebble conversion is converged for particle
  resolution, depending instead on the
  resolution of the mesh. For the $256^3$ model, at the
end of the simulation, only 5998 of the original $\xtimes{7.5}{6}$ superparticles remain, that is,
99.9\% of the pebbles have been converted into protoplanets. The
number was 99.6\% at 200 orbits, 92\% at 100 orbits, and 72\% at 50
orbits. 


\subsection{Varying Stokes number and dust-to-gas ratio}

\Fig{fig:pebble_evolution} tracks the density of pebbles in a
  series of simulations. The fiducial model is the grey curve
  (resolution $256^3$, $Z=10^{-2}$ and ${\rm St}=0.3$).
  The pebble density steadily increases to the Hill density
  ($\approx$ 90 in code units) within less than two orbits after the inclusion
  of pebbles, forming the first protoplanets. As a
  gravitationally bound clump is replaced by a sink particle and the
pebbles of the clump are removed from the simulation, the maximum pebble density
stays constant after reaching Hill density.
Protoplanet formation stops after $\approx$ 150
orbits, after which time the
maximum density drops below the Hill density for the remainder of the
simulation (until 300 orbits). In the $128^3$ simulation (blue curve
in left plot), planet
formation continues intermittently until about 400
orbits. The middle panel shows the fiducial simulation and those of
different Stokes numbers. For St=0.1 planet formation is barely
delayed, from 2 to 10 orbits. For St=0.03, however, even after 150
orbits the pebble concentration has not achieved gravitational
collapse, with the maximum pebble density reaching a plateau at about 10.

The right panel shows the dependency on initial dust-to-gas
  ratio. At $Z=10^{-3}$ planet formation happens nearly as fast
  as in the prior 
St=0.1 case at $Z=10^{-2}$. At $Z=10^{-4}$, the simulation did not
lead to gravitational collapse, although the dust-to-gas ratio did rise
above unity, stabilizing around 3.

To understand why some simulations lead to gravitational
  collapse and some do not, we consider the analytical solution of
  \citet{LyraLin13}. If the pebbles are trapped in a Kida vortex \citep{Kida81}
  in steady state between diffusion, drag force, and stellar vertical gravity
  then the peak pebble density is

\begin{equation}
\frac{{\rm max}(\rho_p)}{\rho_0} = Z \left(\frac{\rm St}{\delta}+1\right)^{1.5}
\label{eq:lyralin13}
\end{equation}  

\noindent where $\delta$ is the dimensionless diffusion coefficient,
similar to the Shakura-Sunyaev $\alpha$ parameter \citep{ShakuraSunyaev73}.

\fig{fig:alpha-rhop} shows the diffusion versus $\rho_p$ relation for
these different runs. The x-axis is a dust density threshold,
$\rho_{p,\rm ts}$, and the y-axis is the diffusion parameter
considering only the mesh cells that have dust density above this
threshold. The diffusion is estimated as simply the vertical
Mach number squared

\begin{equation}
  \alpha_{zz} \equiv \frac{u_z^2}{c_s^2}
\end{equation}

\noindent and the median is taken in both space (all cells above the threshold) and
  time. The time cadence is one orbit.
  For the $Z=10^{-2}$ and $\St=0.3$ run (grey line) the time interval for averaging was 1 to 100 orbits;
  for the $Z=10^{-2}$ and $\St=0.1$ run (orange line) 10 to 32 orbits;
  for $Z=10^{-2}$ and $\St=0.03$ (green line) 75 to 173 orbits;
  $Z=10^{-3}$ and $\St=0.3$  (cyan line) 10 to 26,
  and finally for $Z=10^{-4}$ and $\St=0.3$ (magenta line) 15 to 30
  orbits. We overplot (dotted lines) the steady state solution
  \eqp{eq:lyralin13} for the different combination of parameters. We
  see that the runs for which collapse occurs ($\rho_p$ reaching 90)
  the diffusion is always under the steady-state prediction. For the
  $Z=10^{-2}$ and $\St=0.3$ run, for instance, Hill density should be
  achieved for $\delta\sim\alpha_{zz}$ up to $\xtimes{7}{-4}$, whereas the diffusion is
  about half this number. Conversely, the simulations that do not
  achieve collapse have maximum density consistent with the expected
  for the amount of diffusion measured. For $\St=0.03$ and $Z=10^{-2}$ the
  prediction for $\rho_{p,{\rm max}}\sim 10$ is $\delta
  =\xtimes{3}{-4}$. Indeed, that is approximately the $\alpha_{zz}$
  measured (green line) for that density threshold. Similarly, for
  $\St=0.03$ and $Z=10^{-2}$, the prediction for $\rho_{p,{\rm
      max}}\sim 3$ is again $\delta
  =\xtimes{3}{-4}$, which too is measured (magenta line) for that
  density threshold. This is an interesting agreement between the
  analytical prediction and the simulation, considering that the
  solution of \citet{LyraLin13} does not consider the backreaction of
  the drag force, that becomes significant for dust-to-gas ratios near
  and above unity. Indeed, the fact that there is a decrease of the
  alpha value inside the vortex compared to the outside ($\rho_p \ll
  1$) shows that the pebble concentration depresses the
  diffusion. That the steady-state solution holds at this regime shows
  that the vortex streamlines are not severely modified by the pebble
  concentration, as already shown by \citet{Lyra+18} and \citet{Raettig+21}.  

\section{Conclusion}
\label{sect:conclusion}

In this letter, we investigate the formation of planetary-mass bodies in vortices (formed by
convective overstability), including particle self gravity. The inclusion of self gravity allows us to follow the
gravitational collapse of the particle clumps concentrated in the
vortices. We model the gas for 400 orbits before insertion of the particles
to let the vortex develop. Our main result is that
proto-planets of large size (of the order of the mass of Mars) are formed from
direct gravitational collapse of the pebble clouds, bypassing the
planetesimal stage.

The same conclusion was found over a decade ago by
\citet{Lyra+08a,Lyra+09a,Lyra+09b}, albeit in 2D models in
low-resolution. Here we show that those results were resolved
in the high-mass end, and also not an artifact of the twodimensionality.
Even when we resolve smaller scales, we find that it is protoplanets
that form, and not smaller planetesimals (although we could not
  find convergence with resolution, so planetesimals may form at the
  small mass end.).

In our shearing box 3D simulations, the vortex forms dozens of Mars-mass objects and hundreds of Moon-mass objects,
in agreement with the previous simulations. The
protoplanets rapidly grow, doubling in mass in about 5 orbits. The
mass accretion rate is consistent with those found in
\cite{Lyra+08b,Lyra+09a}, and in agreement with the mass accretion
rate due to pebble accretion \citep{LambrechtsJohansen12}, albeit with a much larger cross section \citep{Cummins+22}. 

In our models, gravitationally bound clumps are
replaced by sink particles, which are not affected by the drag
force. All particles in the cell where the clump is located are
converted into a sink. Unlike calculations where the drag force keeps
influencing the gravitationally bound clump, we see that the
protoplanets drift away from the vortex, which keeps producing
planets. 

We measure the mass function to be well-fit by a power-law $d\ln N/d\ln M=-1.6\pm 0.3$ between lunar 
and Mars mass. Intriguingly, even though our model does not resolve the streaming instability,
a power-law index of -1.6 was also found by \citet{Johansen+15} (for sufficiently small masses in their mass function) and 
\citet{Simon+16} in streaming instability simulations. That
  different processes (streaming instability and vortex trapping)
  lead to the same slope for the mass function might suggest a
  universality in the outcome of planet formation, irrespective of formation process. However, more recent works \citep[e.g.][]{Abod+19,Li+19,Gole+20}
  find slightly shallower power-laws, when exponential truncation of
  the power law is used. In these works, such shallower power-laws and an exponential truncation was a better fit than the -1.6 slope. In our vortex
  models, we do not see a clear need for tapering, as opposed to the mass function of the streaming instability \cite[e.g.,][]{Schafer+17,Abod+19}.
  The difference seen could also be due to resolution, since we attain only $128/H$ in the highest resolution we ran, compared to $320/H$ in the lowest resolution modeled by
    \citet{Johansen+15,Simon+16}, and $2560/H$ in their highest. A model that resolves streaming instability and vortex trapping in the same simulation
    will be necessary to settle the question.
  
Our results show that vortices are efficient formation sites of
Mars-mass planetary embryos. In the present work, Pluto-mass
objects are still formed via direct gravitational collapse.
However, even the smallest objects formed are still
  about 10 times the Toomre mass. Clearly, the Toomre
  mass is not the characteristic mass of the problem \citep[see also][]{Abod+19}. It remains to be explored whether streaming instability would operate inside a vortex, which could potentially lead to smaller masses at formation. This would necessitate even higher resolution models, which we will explore in future work. 


\section*{Acknowledgments}
  
We thank Anders Johansen, Hubert
  Klahr, Debanjan Sengupta, and Daniel Carrera for insightful discussions, and the
  anonymous referee for a helpful report. We acknowledge support from the NASA 
Theoretical and Computational Astrophysical Networks (TCAN) via grant
\#80NSSC21K0497. WL is further supported by grants
\#80NSSC22K1419 from the NASA Emerging Worlds program, NSF via grant
AST-2007422, and CCY acknowledges the support from NASA via the
Astrophysics Theory Program (grant \#80NSSC24K0133) and the Emerging
Worlds program (grant \#80NSSC23K0653).

The simulations presented in this paper utilized the Stampede cluster of the Texas Advanced Computing Center (TACC) at The University of Texas at Austin, through XSEDE/Access grant TG-AST140014, and the Discovery cluster at New Mexico State University \citep{TrecakovVonWolff21}. This work utilized resources from the New Mexico State University High Performance Computing Group, which is directly supported by the National Science Foundation (OAC-2019000), the Student Technology Advisory Committee, and New Mexico State University and benefits from inclusion in various grants (DoD ARO-W911NF1810454; NSF EPSCoR OIA-1757207; Partnership for the Advancement of Cancer Research, supported in part by NCI grants U54 CA132383 (NMSU)). This research was made possible by the open-source projects \texttt{jupyter}\ \citep{Jupyter}, \texttt{IPython}
\citep{IPython}, \texttt{matplotlib} \citep{matplotlib1, matplotlib2}, \texttt{NumPy} \citep{numpy}, \texttt{SymPy} \citep{sympy}, and \texttt{AstroPy} \citep{astropy}. 



\end{document}